\documentclass[fleqn,12pt]{article}
\usepackage{epsfig}

\newcommand\pubnumberc{CERN--TH/2001--018}
\newcommand\pubnumberk{KA--TP--2--2001}
\newcommand\pubdate{\today}
\newcommand\hepnumber{hep-ph/0101260}

\def\csumb{$^a$ DESY Theorie, Notkestr.\ 85, D-22603 Hamburg, Germany \\[0.2cm]
$^b$ HET, Physics Department, Brookhaven Nat.\ Lab.,
    Upton, NY 11973, USA \\[0.2cm]
$^c$ Institut f\"ur Theoretische Physik,
Universit\"at Karlsruhe, \\ D-76128 Karlsruhe, Germany\\[0.2cm] 
$^d$ CERN, TH Division, CH-1211 Geneva 23, Switzerland}
\def\support{\footnote{Work supported by the
Deutsche Forschungsgemeinschaft and by the European Union.}} 

\def\Title#1{\begin{center} {\Large\bf #1 } \end{center}}
\def\Author#1{\begin{center}{ \sc #1} \end{center}}
\def\Address#1{\begin{center}{ \it #1} \end{center}}

\newcommand\pubblock{\rightline{\begin{tabular}{l} \pubnumberc\\
                                                   \pubnumberk\\
         \pubdate\\ \hepnumber \end{tabular}}}
\newenvironment{Abstract}{\begin{quotation}  }{\end{quotation}}
\newenvironment{Presented}{\begin{quotation} \begin{center} 
             Presented by {\sc W. Hollik} at the\end{center}
      \begin{center}\begin{large}}{\end{large}\end{center} \end{quotation}}
\def\Acknowledgments{\bigskip  \bigskip \begin{center}
          \large\bf Acknowledgments\end{center}}

\makeatletter
\def\section{\@startsection{section}{0}{\z@}{5.5ex plus .5ex minus
 1.5ex}{2.3ex plus .2ex}{\large\bf}}
\def\subsection{\@startsection{subsection}{1}{\z@}{3.5ex plus .5ex minus
 1.5ex}{1.3ex plus .2ex}{\normalsize\bf}}
\def\subsubsection{\@startsection{subsubsection}{2}{\z@}{-3.5ex plus
-1ex minus  -.2ex}{2.3ex plus .2ex}{\normalsize\sl}}

\renewcommand{\@makecaption}[2]{%
   \vskip 10pt
   \setbox\@tempboxa\hbox{\small #1: #2}
   \ifdim \wd\@tempboxa >\hsize     
       \small #1: #2\par          
     \else                        
       \hbox to\hsize{\hfil\box\@tempboxa\hfil}
   \fi}

 \def\citenum#1{{\def\@cite##1##2{##1}\cite{#1}}}
 
\newcount\@tempcntc
\def\@citex[#1]#2{\if@filesw\immediate\write\@auxout{\string\citation{#2}}\fi
  \@tempcnta\z@\@tempcntb\m@ne\def\@citea{}\@cite{\@for\@citeb:=#2\do
    {\@ifundefined
       {b@\@citeb}{\@citeo\@tempcntb\m@ne\@citea\def\@citea{,}{\bf ?}\@warning
       {Citation `\@citeb' on page \thepage \space undefined}}%
    {\setbox\z@\hbox{\global\@tempcntc0\csname b@\@citeb\endcsname\relax}%
     \ifnum\@tempcntc=\z@ \@citeo\@tempcntb\m@ne
       \@citea\def\@citea{,}\hbox{\csname b@\@citeb\endcsname}%
     \else
      \advance\@tempcntb\@ne
      \ifnum\@tempcntb=\@tempcntc
      \else\advance\@tempcntb\m@ne\@citeo
      \@tempcnta\@tempcntc\@tempcntb\@tempcntc\fi\fi}}\@citeo}{#1}}
\def\@citeo{\ifnum\@tempcnta>\@tempcntb\else\@citea\def\@citea{,}%
  \ifnum\@tempcnta=\@tempcntb\the\@tempcnta\else
  {\advance\@tempcnta\@ne\ifnum\@tempcnta=\@tempcntb \else\def\@citea{--}\fi
    \advance\@tempcnta\m@ne\the\@tempcnta\@citea\the\@tempcntb}\fi\fi}
\makeatother

\def\beq{\begin{equation}}
\def\eeq#1{\label{#1}\end{equation}}
\def\eeqn{\end{equation}}

\newenvironment{Eqnarray}%
   {\arraycolsep 0.14em\begin{eqnarray}}{\end{eqnarray}}
\def\beqa{\begin{Eqnarray}}
\def\eeqa#1{\label{#1}\end{Eqnarray}}
\def\eeqan{\end{Eqnarray}}

\let\bar=\overbar

\def\Dslash{\not{\hbox{\kern-4pt $D$}}}
\def\dslash{\not{\hbox{\kern-2pt $\del$}}}

\def\mt{m_t}

\def\GF{G_F}

\def\msb{{\bar{\ssstyle M \kern -1pt S}}}

\def\lsim{\mathrel{\raise.3ex\hbox{$<$\kern-.75em\lower1ex\hbox{$\sim$}}}}
\def\gsim{\mathrel{\raise.3ex\hbox{$>$\kern-.75em\lower1ex\hbox{$\sim$}}}}

\def\al{\alpha}

\def\de{\delta}

\def\De{\Delta}
\def\Re{\mathrm{Re}}
\def\Im{\mathrm{Im}}

\newcommand{\stxt}[1]{\mbox{\scriptsize #1}}

\newcommand{\MW}{M_W}
\newcommand{\MZ}{M_Z}
\newcommand{\MH}{M_H}
\newcommand{\OMT}[1]{\overline{M}_{\hspace{-.3ex}{#1}}}
\newcommand{\OMP}{\overline{M}}
\newcommand{\OGP}{\overline{\Gamma}}

\newcommand{\Mt}{m_t}
\newcommand{\sw}{s_{\mbox{\tiny{W}}}}
\newcommand{\cw}{c_{\mbox{\tiny{W}}}}
\newcommand{\as}{\alpha_s}
\newcommand{\alps}{\alpha_s}

\newcommand{\msbar}{$\overline{MS}$}

\newcommand{\rt}[1]{\left(#1\right)^{\frac{1}{2}}}
\newcommand{\irt}[1]{\left(#1\right)^{-\frac{1}{2}}}

\newcommand{\xia}{\xi_1^{\gamma}}
\newcommand{\xiaz}{\xi^{\gamma Z}}
\newcommand{\xiza}{\xi^{Z\gamma}}
\newcommand{\xiz}{\xi_1^Z}
\newcommand{\xizz}{\xi_2^Z}
\newcommand{\xiw}{\xi_1^W}
\newcommand{\xiww}{\xi_2^W}

\newcommand{\gev}{\unskip\,\mathrm{GeV}}

\newcommand{\fea}{{\em FeynArts}}

\newcommand{\eqref}[1]{(\ref{#1})}

\begin{document}
\begin{titlepage}
\pubblock

\vfill
\def\thefootnote{\fnsymbol{footnote}}
\Title{Two-loop electroweak contributions 
to $\Delta r$ \support}  
\vfill
\Author{A. Freitas$^a$, S. Heinemeyer$^b$, 
        W. Hollik$^c$, W. Walter$^c$, 
        G. Weiglein$^d$} 
\Address{\csumb}
\vfill
\begin{Abstract}
A review is given on the quantum correction
$\Delta r$ in the $W$--$Z$ mass correlation at
the electroweak two-loop level, as derived from 
the calculation of the muon lifetime in the Standard Model. 
Exact results for $\Delta r$ and the $W$-mass prediction
including ${\mathcal{O}}(\alpha^2)$ corrections with fermion loops are
presented and compared with previous results of a next-to-leading order
expansion in the top-quark mass. 
\end{Abstract}
\vfill
\begin{Presented}
5th International Symposium on Radiative Corrections \\ 
(RADCOR--2000) \\[4pt]
Carmel CA, USA, 11--15 September, 2000
\end{Presented}
\vfill
\end{titlepage}
\def\thefootnote{\arabic{footnote}}
\setcounter{footnote}{0}

\section{Introduction}

The interdependence between the $W$-boson mass, $\MW$, and the $Z$-boson
mass, $\MZ$, with the help of the Fermi constant
$\GF$ and the fine structure constant $\al$ 
is one of the most important relations for testing the
electroweak Standard Model (SM) with high precision.
At present, the world-average for the $W$-boson mass is 
$\MW^{\mathrm{exp}} = 80.434 \pm 0.037$~GeV~\cite{mori00}.
The experimental precision on $\MW$
will be further improved with the data taken at LEP2 in their final
analysis, at the upgraded Tevatron and at the LHC, where 
an error of $\de\MW = 15$~MeV can be expected~\cite{lhctdr}. At a 
high-luminosity linear collider running in a low-energy mode at the 
$W^+W^-$ threshold, a reduction 
of the experimental error down to 
$\de\MW = 6$~MeV may be feasible~\cite{gigazMW}. This offers the
prospect for highly sensitive tests of the electroweak
theory~\cite{gigaztests}, provided that the accuracy of the theoretical
prediction matches the experimental precision.
The basic physical quantity for the $\MW$--$\MZ$ correlation 
is the muon lifetime $\tau_{\mu}$,
which defines  the Fermi constant $\GF$
according to
\beq
\frac{1}{\tau_{\mu}} = \frac{\GF^2 \, m_\mu^5}{192 \pi^3} \;
F\left(\frac{m_e^2}{m_\mu^2}\right)
\left(1 + \frac{3}{5} \frac{m_\mu^2}{\MW^2} \right) 
\left(1 + \Delta_{\rm QED} \right) ,
\label{eq:fermi}
\end{equation}
with $F(x) = 1 - 8 x - 12 x^2 \ln x + 8 x^3 - x^4$. By convention, the
QED corrections within the Fermi Model, $\Delta_{\rm QED}$, 
are included in this defining equation for $\GF$. 
The one-loop result for 
$\Delta_{\rm QED}$~\cite{delqol}, 
which has already been known for several
decades, has recently been supplemented by the two-loop
correction~\cite{delqtl}, yielding 
\beq 
\Delta_{\rm QED} = 1 - 1.81\, \frac{\alpha(m_\mu)}{\pi}
                       +6.7 \left( \frac{\alpha(m_\mu)}{\pi} \right)^2 ,
 \quad \mbox{with} \quad
\alpha(m_\mu) \simeq \frac{1}{135.90} \;\; .
\end{equation}
The tree-level $W$-propagator effect giving rise to the
(numerically insignificant) term $3 m_\mu^2/(5 \MW^2)$ in 
\eqref{eq:fermi}, is conventionally also included in the definition of
$\GF$, although not part of the Fermi Model prediction.
From the precisely measured muon-decay width the value
\cite{C:pdg} $\GF = (1.16637 \pm 0.00001)\, 10^{-5} \mbox{ GeV}^{-2}$
for the Fermi constant is derived.

Calculating 
the muon lifetime within the SM and comparing the SM result 
with~\eqref{eq:fermi} yields the relation
\beq
\MW^2 \left(1 - \frac{\MW^2}{\MZ^2}\right) = 
\frac{\pi \al}{\sqrt{2} \GF} \left(1 + \De r\right),
\label{eq:delr}
\end{equation}
where the radiative corrections are summarized 
in the quantity $\De r$, first calculated in~\cite{sirlin}
at one-loop.
This relation can be used for deriving the prediction of $\MW$ within 
the SM (or possible extensions), to be confronted with the experimental 
result for $\MW$.

The one-loop result for $\De r$ within the SM
can be decomposed as follows (with the notation
$\sw^2 = 1 - \MW^2/\MZ^2, \, \cw^2 = 1- \sw^2$),
\beq
\De r^{(\al)} = \De \al - \frac{\cw^2}{\sw^2} \De\rho + 
\De r_{\mathrm{rem}}(\MH),
\label{eq:delrol}
\end{equation}
exhibiting the leading fermion-loop contributions $\De \al$ and $\De\rho$,
which originate from the charge and mixing-angle renormalization;
the remainder part $\De r_{\mathrm{rem}}$ contains 
in particular the dependence on the Higgs-boson mass, $\MH$. 
The QED-induced shift $\De \al$
in the fine structure constant contains 
large logarithms of light-fermion masses. 
The leading contribution to the $\rho$~parameter from
the top/bottom weak-isospin doublet, $\De\rho$, gives rise to a term
with a quadratic dependence on the top-quark mass, $\mt$~\cite{velt}.

Beyond the one-loop order, resummations of the leading one-loop
contributions $\De\al$ and $\De\rho$ are known~\cite{resum}.
They correctly take into account the terms of the form 
$(\De\al)^2$, $(\De\rho)^2$, $(\De\al\De\rho)$, and
$(\De\al\De r_{\mathrm{rem}})$ at the two-loop level and the leading
powers in $\De\al$ to all orders. 

QCD corrections to $\De r$ are known at ${\cal O}(\al
\alps)$~\cite{qcd2} and ${\cal O}(\al \alps^2)$~\cite{qcd3}. 
Concerning the electroweak two-loop contributions, only 
partial results are available up to now. 
Approximative calculations were performed
based on expansions for
asymptotically large values of 
$\MH$~\cite{ewmh2} and $\mt$~\cite{ewmt4,ewmt2,rho3}.
The terms derived by expanding in the top-quark
mass of ${\cal O}(\GF^2 \Mt^4)$~\cite{ewmt4} and 
${\cal O}(\GF^2 \Mt^2 \MZ^2)$~\cite{ewmt2} were found to be numerically
sizeable. The ${\cal O}(\GF^2 \Mt^2 \MZ^2)$ term, involving three different 
mass scales, has been obtained by two separate expansions in the regions
$\MW, \MZ, \MH \ll \mt$ and $\MW, \MZ \ll \mt, \MH$ and by an interpolation
between the two expansions. This formally next-to-leading order term
turned out to be of a magnitude similar to that of
the formally leading term of ${\cal O}(\GF^2 \Mt^4)$, entering with
the same sign. Its inclusion (both for $\MW$ and the effective weak
mixing angle) had important consequences on the indirect constraints
on the Higgs-boson mass derived from the SM fit to the precision data.

A more complete calculation of electroweak two-loop
effects is hence desirable.
As a first step in this direction, exact results were derived 
for the Higgs-mass dependence of the fermionic two-loop 
corrections to the precision observables~\cite{ewmhdep}. They have been
compared with the results of expanding up to ${\cal O}(\GF^2 \Mt^2
\MZ^2)$~\cite{ewmt2}, specifically analysing the effects of the $\mt$
expansion, and good agreement has been found~\cite{gsw}.
Beyond the two-loop order, complete results for the pure fermion-loop 
corrections ({\it i.e.}~contributions containing $n$ fermion loops at 
$n$-loop order) have recently been obtained up to four-loop order~\cite{floops}.
These results contain in particular the contributions of the leading
powers in $\De\al$ as well as the ones in $\De\rho$ and the mixed
terms.

\begin{figure}[ht]
\vspace{1em}
\begin{center}
\psfig{figure=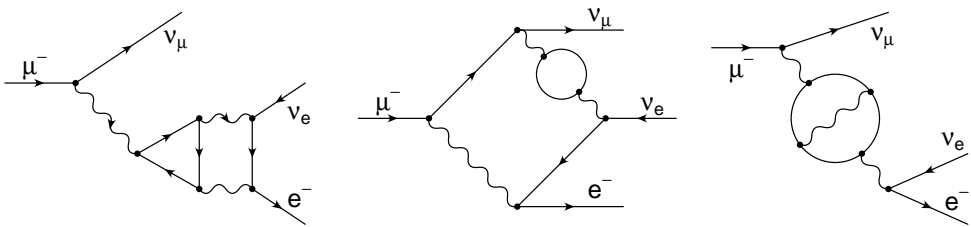,width=13cm}
\vspace{-1em}
\end{center}
\caption[]{{\small
Examples for various types of fermionic two-loop diagrams contributing to muon
decay.
\label{fig:diags}
}}
\end{figure}

In this talk the complete 
fermionic electroweak two-loop corrections to $\De r$
are discussed, as calculated exactly 
without an expansion in the top-quark or the
Higgs-boson mass \cite{plb-paper}.
These are all two-loop diagrams contributing to 
the muon-decay amplitude and
containing at least one closed fermion loop (except the pure QED
corrections already contained in the Fermi model result, according 
to~\eqref{eq:fermi}). 
Fig.~{\ref{fig:diags} displays some typical examples}.
The considered class of diagrams includes the
potentially large corrections both from the top/bottom doublet and 
from contributions proportional to $N_{lf}$ and $N_{lf}^2$, 
where $N_{lf}$ is the number of light fermions (a partial result for 
the light-fermion contributions is given in \cite{stuartlf}). The 
results presented here improve on the previous results of an expansion in $\mt$
up to next-to-leading order~\cite{ewmt2} in containing the full
dependence on $\mt$ as well as the complete light-fermion contributions
at the two-loop order, while in \cite{ewmt2} higher-order corrections from
light fermions have only been taken into account via a resummation of
the one-loop light-fermion contribution.

\section{Outline of the calculation}

Since all possibly infrared (IR) divergent photonic corrections are already
contained in the definition \eqref{eq:fermi} of the Fermi constant $\GF$ and mass
singularities are absorbed in the running of the electromagnetic coupling, 
$\MW$ represents the scale for the electroweak corrections in $\Delta
r$. Therefore it is possible to neglect all fermion masses except the top-quark 
mass and the momenta of the external leptons so that the Feynman diagrams
for muon decay reduce to vacuum diagrams.

All QED contributions to the Fermi Model have to be excluded in the computation
of $\Delta r$ since they have already been
separated off in the definition of $\GF$, see eq.~\eqref{eq:fermi}. 
Apart from the one-loop contributions, this comprises
two-loop QED corrections
and mixed contributions of QED and weak corrections of each one-loop order,
which have to be removed
from $\Delta r^{(\alpha^2)}$. For fermionic two-loop diagrams it is 
possible to find a one-to-one correspondence between QED graphs in Fermi-Model
and SM contributions.

After extracting the IR-divergent QED corrections,
the generic diagrams contributing to the muon-decay amplitude
can be reduced to vacuum-type diagrams, 
since the masses of the external particles  and
the momentum transfer are negligible.
The renormalization of is performed in
the on-shell scheme.
Thus, the mass-renormalization of the gauge bosons
requires the evaluation of two-loop two-point functions with non-zero 
external momentum, which is more involved from a technical point of view
regarding the tensor structure and the evaluation of the scalar
integrals. 
This complication cannot be avoided by 
performing the calculation within another renormalization scheme, 
{\it e.g.}~the \msbar\ scheme, 
since ultimately one is interested in the relation
between the physical parameters $\MW$, $\MZ$, $\al$, $\GF$,
where the two-point functions for non-zero momenta enter.

All diagrams and amplitudes for the decay and counterterm contributions have
been generated with the program \emph{FeynArts 2.2} \cite{fea}. The
amplitudes are algebraically reduced by means of a general tensor-integral
decomposition for two-loop two-point functions with the program \emph{TwoCalc}
\cite{two}, leading to a fixed set of standard scalar integrals. Analytic
expressions are known for the scalar one-loop \cite{HV:oneloop} and two-loop
\cite{davtausk} vacuum integrals, whereas the two-loop self-energy diagrams can
be evaluated numerically by means of one-dimensional integral
representations \cite{intnum}.

In order to apply an additional check the calculations were performed within a
covariant $R_\xi$ gauge, with individual gauge parameters $\xi_i$
for each gauge boson. It has been explicitly checked at the
algebraic level that the gauge-parameter dependence of the final result drops
out.

At the subloop level, also the Faddeev-Popov ghost sector
has to be renormalized.
The gauge-fixing part of the Lagrangian, in terms of the gauge fields
$A^\mu$, $Z^\mu$, $W^{\pm\mu}$ and the
unphysical Higgs scalars $\chi$, $\phi^\pm$ given by
\[
{\mathcal{L}}_{\stxt{gf}} = -\frac{1}{2}
  \Big[(F^\gamma)^2+(F^{{Z}})^2+F^+F^-+F^-F^+\Big], 
    \quad \quad \quad \mbox{with}
\]
\vspace{-2.5ex}
\begin{equation}
F^\gamma =\irt{\xia} \partial_\mu A^\mu +\frac{\xiaz}{2} \partial_\mu Z^\mu,
\label{gaugefix}
\end{equation}
\vspace{-2ex}
\[
F^{{Z}} = \irt{\xiz} \partial_\mu Z^\mu 
        + \frac{\xiza}{2} \partial_\mu A^\mu
        - \rt{\xizz} \MZ \, \chi,
\]
\vspace{-2.5ex}
\[
F^\pm =\irt{\xiw}\partial_\mu W^{\pm\mu} \,\mp\,i\rt{\xiww} \MW
 \, \phi^\pm \, ,
\]
does not need renormalization.
Accordingly, one can either introduce the gauge-fixing term 
after renormalization or 
renormalize the gauge parameters in
such a way that they compensate the renormalization of the fields and masses. 
Both ensure that no counterterms arise from the gauge-fixing sector but they 
differ in the treatment of the ghost Lagrangian, 
which is given by the variation of the
functionals $F^a$ under infinitesimal gauge transformations
$\delta\theta_b$,
\begin{equation}
{\mathcal{L}}_{\stxt{FP}} = \sum_{a,b = \gamma, Z, \pm} \bar{u}^a \,
  \frac{\delta F^a}{\delta \theta^b} \, u^b. \label{FP}
\end{equation}
In the latter case, which was applied in our work for simplification
of the automatized treatment, 
additional counterterm contributions for the ghost sector arise from the
gauge-parameter renormalization.
The parameters $\xi^a_i$ in \eqref{gaugefix} are renormalized 
such that their counterterms  $\delta\xi^a_i$ exactly cancel the
contributions from the renormalization of the fields and masses and that the
renormalized gauge parameters comply with the $R_\xi$ gauge.

\section{On the ${\mbox \boldmath \gamma_5}$--problem}

In four dimensions the algebra of the $\gamma_5$--matrix is defined by the two
relations
\begin{eqnarray}
& & \bigl\{ \gamma_5, \gamma_\alpha \bigr\} = 0 \qquad \mbox{for} \qquad \alpha =
  1, \dots, 4 \label{anticomm} \\[1ex]
& & \mbox{Tr} \bigl\{ \gamma_5 \gamma^\mu\gamma^\nu\gamma^\rho\gamma^\sigma \bigr\}
  = 4i \epsilon^{\mu\nu\rho\sigma}\mbox{.} \label{trg4}
\end{eqnarray}
It is impossible to translate both relations simultaneously into $D
\neq 4$ dimensions without encountering inconsistencies \cite{ga5HV}.

A certain treatment of $\gamma_5$ might break symmetries, {\it i.e.}~violate
Slavnov-Taylor (ST) identities which would have to be restored with extra
counterterms. Even after this procedure a residual scheme dependence can
persist which is associated with $\epsilon$-tensor expressions originating from
the treatment of \eqref{trg4}. Such expressions cannot be canceled by
counterterms. If they broke ST identities this would give rise to anomalies.

't Hooft and Veltman \cite{ga5HV} suggested a consistent scheme,
formalized by Breitenlohner and Maison \cite{ga5BM}, as a separation of
the first four and the remaining dimensions of the $\gamma$-Matrices
(HVBM-scheme). It has been shown \cite{SMren} that the SM with HVBM
regularization is anomaly-free and renormalizable. This shows that
$\epsilon$-tensor terms do not get merged with divergences.

The naively anti-commuting scheme, which is widely used for one-loop
calculations, extends the rule \eqref{anticomm} to $D$ dimensions but abandons
\eqref{trg4},
\begin{eqnarray} & & 
\bigl\{ \gamma_5, \gamma_\alpha \bigr\} = 0 \qquad \mbox{for} \qquad \alpha =
  1, \dots, D \\[1ex] & & 
\mbox{Tr} \bigl\{ \gamma_5 \gamma^\mu\gamma^\nu\gamma^\rho\gamma^\sigma \bigr\}
  = 0.
\end{eqnarray}
This scheme is unambiguous but does not reproduce the four-dimensional case.

\begin{figure}[tb]
\centerline{
\psfig{figure=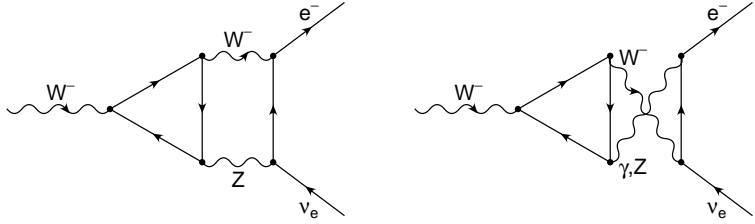,width=10cm}
}
\caption{{\small
Charged-current vertex diagrams with fermion-triangle subgraphs.}}
\label{fig:CCvert}
\end{figure}

In the SM, particularly triangle diagrams 
(like the ones in Figure~\ref{fig:CCvert})
containing chiral couplings are sensitive to the $\gamma_5$--problem. In our
context, the one-loop triangle diagrams have been explicitly calculated in both
schemes. While the naive scheme immediately respects all ST
identities the HVBM scheme requires the introduction of additional finite
counterterms. Even after this procedure finite differences remain between the
results of the two schemes, showing that the naive scheme is
inapplicable in this case.

In the calculation of $\Delta r$ triangle diagrams appear as subloops of
two-loop charged current (CC) vertex diagrams (Fig.~\ref{fig:CCvert}). One
finds that for the difference terms between both schemes this loop can be
evaluated in four dimensions without further difficulties. This can be
explained by the fact that renormalizability forbids divergent contributions to
$\epsilon$-tensor terms from higher loops in the HVBM scheme. The
$\epsilon$-tensor contributions from the triangle subgraph in the HVBM scheme
meet a second $\epsilon$-tensor term from the outer fermion lines in 
Fig.~\ref{fig:CCvert}, 
thereby resulting in a non-zero contribution to $\Delta r$.

Computations in the HVBM scheme can in general
get very tedious because of the necessity
of additional counterterms. 
For our specific problem, however, it is possible to apply  
another, simplified,  method. One
can consider a ``mixed'' scheme that uses both relations \eqref{anticomm} and
\eqref{trg4} in $D$ dimensions, despite their mathematical inconsistency, 
to evaluate the one-loop triangle subgraphs.
These results 
immediately respect all ST identities.
As checked explicitly, they differ from 
the HVBM results (with the appropriate counterterms to restore the
ST identities) 
only by terms of ${\mathcal{O}}(D-4)$,
\begin{equation}
\Gamma^{\stxt{HVBM}}_{\Delta (1)} =
\Gamma^{\stxt{mix}}_{\Delta (1)} + {\mathcal{O}}(D-4).
\end{equation}
Inserting the one-loop expressions into the the two-loop diagrams  
one finds that the second loop integration gives a finite result
and hence can be performed in four dimensions yielding
\begin{equation}
\Gamma^{\stxt{HVBM}}_{\stxt{CC(2)}} =
\Gamma^{\stxt{mix}}_{\stxt{CC(2)}} + {\mathcal{O}}(D-4).
\end{equation}
Thus the mixed scheme can serve in this case  
as a technically easy prescription for the correct calculation of the
CC two-loop contributions. Practical ways of treating $\gamma_5$ 
in higher-order calculations are also discussed  in~\cite{jeger}.

\section{Two-loop renormalization}

For the determination of the one-loop counterterms  and renormalization
constants the conventions of~\cite{Dehabil} are adopted. 
Two-loop renormalization constants 
enter via the counterterms for the transverse $W$ propagator and the charged
current vertex (the counterterms for the transverse $Z$ propagator
are analogous):
\begin{eqnarray}
\hspace{-4ex}
\left[
\mbox{\raisebox{-1mm}{\psfig{figure=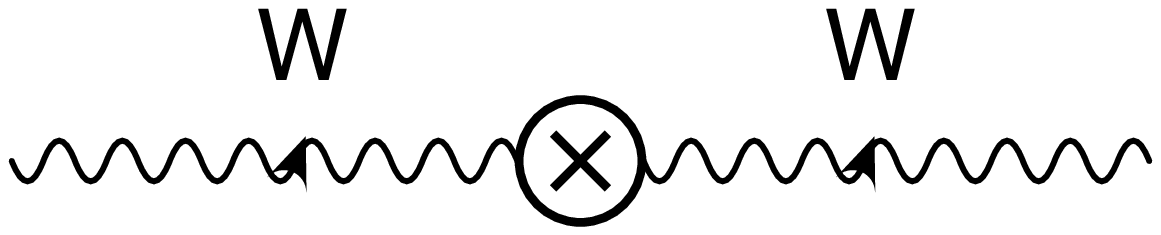,width=30mm}}}
\right]_{\stxt{T}}
 &=&
 \delta Z_{(2)}^{{W}}(k^2-\MW^2) - \delta M^2_{W(2)} -
        \delta Z_{(1)}^{{W}} \delta M^2_{{W(1)}}, \label{w-ct} \\[3ex]
\hspace{-4ex}
\mbox{\raisebox{-8mm}{\psfig{figure=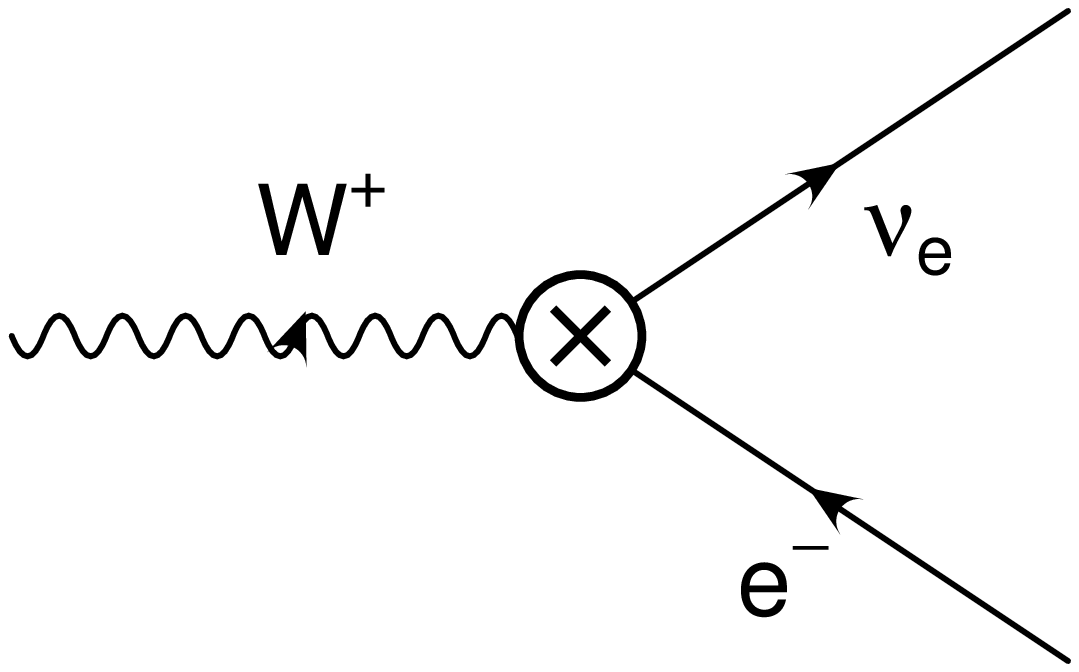,width=25mm}}}
 &=& i \frac{e}{\sqrt{2} \sw} \, \gamma_\mu \omega_-
  \biggl[ \delta Z_{e(2)}
  - \frac{\delta s_{\mbox{\stxt{W(2)}}}}{s_{\stxt{W}}}
  +\frac{1}{2} \left(\delta Z_{(2)}^{e\stxt{L}}+\delta Z_{(2)}^{{W}} +
        \delta Z_{(2)}^{\nu \stxt{L}} \right) \nonumber  \\[-1ex]
 &&+ \; \mbox{(1-loop renormalization constants)} \biggr]. 
\end{eqnarray}
$\delta Z^{W,e\stxt{L},\nu\stxt{L}}$ 
denote the field-renormalization constants, 
$\delta M^2_{W,Z}$ the $W$- and $Z$-mass counterterms, 
and $\delta Z_e$ denotes the 
charge-renormalization constant. The lower indices 
in parentheses indicate the loop order. The mixing-angle  
counterterm $\delta s_{\stxt{W(2)}}$ 
can be derived from 
the gauge-boson mass counterterms. 
The two-loop contributions always include the subloop
renormalization.


The on-shell masses are defined as the position of the propagator poles. 
Starting at the two-loop level, 
it has to be taken into account that there is a difference
between the definition of the mass $\widetilde{M}^2$ as the 
pole of the real part of the
(transverse) propagator,
\begin{equation}
\Re \left\{(D_{\stxt{T}})^{-1}({\widetilde{M}}^2) \right\}
 = 0,
\end{equation}
and the real part $\overline{M}^2$ of the complex pole,
\begin{equation}
\label{complex}
(D_{\stxt{T}})^{-1}({\mathcal{M}}^2) = 0, \qquad
{\mathcal{M}}^2 = \OMP^2  - i \OMP\, \OGP.
\end{equation}
The imaginary part of the complex pole is associated with the width
$\overline{\Gamma}$. The defining condition (\ref{complex})
yields for the $W$-mass counterterm
\begin{equation}
\delta \OMT{W(2)}^2 = \Re \bigl\{
  \Sigma_{\stxt{T(2)}}^{{W}}(\OMT{W}^2) \bigr\}
 -  \delta Z_{(1)}^{{W}} \delta \OMT{W(1)}^2
 + \Im\bigl\{\Sigma_{\stxt{T}(1)}^{{W}^/}(\OMT{W}^2)\bigr\}
     \:
   \Im\bigl\{\Sigma_{\stxt{T}(1)}^{{W}}(\OMT{W}^2)\bigr\}, \label{MWct}
     \nonumber
\end{equation}
whereas for the real-pole definition the last term of 
eq.~\eqref{MWct} is missing. 
$\Sigma_{\stxt{T}}^{{W}}$
denotes the transverse $W$ self-energy and 
$\Sigma_{\stxt{T}}^{{W}^/}$  
its momentum derivative. 
Similar expressions hold for the $Z$ boson.

The $W$ and $Z$ mass countertermss determine 
the two-loop counterterm for the mixing angle, $\delta
s_{\stxt{W(2)}}$, which has to be gauge invariant since $\sw$ is an observable
quantity. With the use of a general $R_\xi$ gauge it has
explicitly been checked that $\delta s_{\stxt{W(2)}}$
is gauge-parameter independent for the complex-pole mass definition,
whereas the real-pole definition leads to a gauge dependent
$\delta s_{\stxt{W(2)}}$.
This is in accordance with the expectation from $S$-matrix theory
\cite{unstab}, where
the complex pole represents a gauge-invariant mass definition.

It should be noted that the mass definition via the complex pole corresponds to
a Breit-Wigner parameterization of the resonance shape with a constant width.
For the experimental determination of the gauge-boson masses, however,
a Breit-Wigner ansatz with a running width is used. This has to be 
accounted for
by a shift of the values for the complex pole masses 
\cite{massshift},
\begin{equation}
\OMP = M - \frac{\Gamma^2}{2 M}.
\end{equation}
which yields the relations
\begin{eqnarray}
\begin{array}{lcl}
 \OMT{Z} & = & \MZ - 34.1 \mbox{ MeV,}  \\
 \OMT{W} & = & \MW - 27.4\; (27.0) \mbox{ MeV} \quad 
  \mbox{ for } \quad \MW = 80.4\; (80.2) \mbox{ GeV.}
\end{array}
\end{eqnarray}
For $\MZ$ and $\Gamma_{{Z}}$ the experimental numbers are taken.
The $W$ mass is a calculated quantity, 
and therefore also a theoretical value for the $W$-boson
width should be applied here.
The results above are obtained from the approximate, but sufficiently
accurate expression for the $W$ width, 
\begin{equation}
\Gamma_{{W}} = 3 \frac{\GF \MW^3}{2 \sqrt{2} \pi} \,
 \left( 1 + \frac{2 \as}{3 \pi} \right) .
\end{equation}

\section{Results}

\begin{figure}[tb]
\centerline{
\psfig{figure=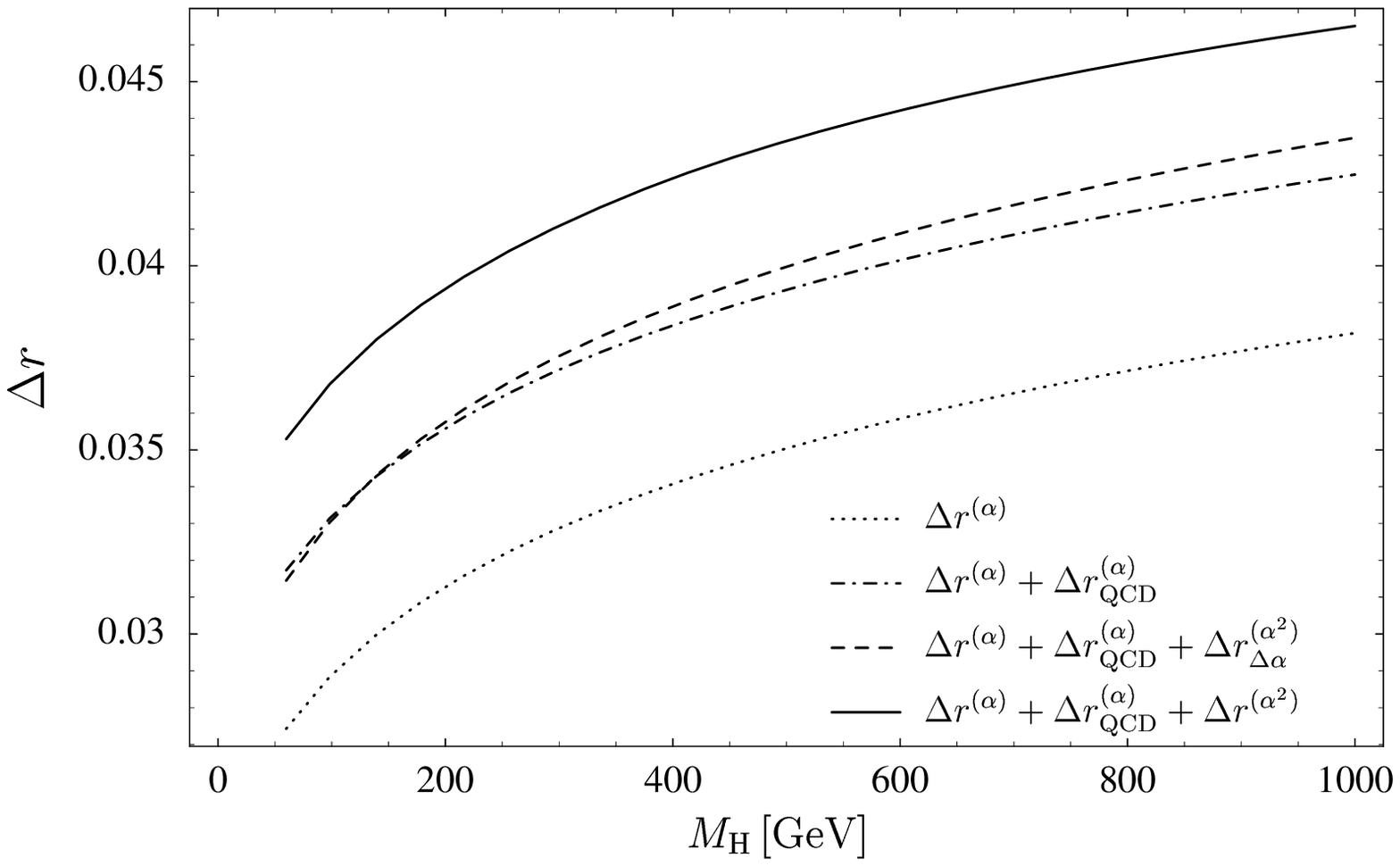,width=14cm}
}
\caption[]{{\small
Various stages of $\De r$, as a function of $\MH$.
The one-loop contribution, $\De r^{(\al)}$, is supplemented by the
two-loop and three-loop QCD corrections, 
$\De r^{(\al)}_{\mathrm{QCD}} \equiv \De r^{(\al\alps)} + \De
r^{(\al\alps^2)}$,
and the fermionic electroweak two-loop contributions,
$\De r^{(\al^2)} \equiv \De r^{(N_{\mathrm{f}} \al^2)} +
\De r^{(N_{\mathrm{f}}^2 \al^2)}$.
For comparison, the effect of the two-loop corrections induced by a
resummation of $\De\al$, $\De r^{(\al^2)}_{\De\al}$, is shown
separately.
}}
\label{fig:deltar}
\end{figure}

In the previous sections the characteristics of the calculation of electroweak
two-loop contributions to $\Delta r$ have been pointed out. Combining the
fermionic ${\mathcal{O}}(\alpha^2)$ contributions with the one-loop and the QCD
corrections yields the total result
\begin{eqnarray}
\Delta r = \Delta r^{(\alpha)} + \Delta r^{(\alpha\alpha_s)} +
  \Delta r^{(\alpha\alpha_s^2)} +
  \Delta r^{(N_f \alpha^2)} + \Delta r^{(N_f^2 \alpha^2)}. \label{drtotal}
\end{eqnarray}
Here $N_f, N_f^2$ symbolize one and two fermionic loops, respectively. 
Fig.~\ref{fig:deltar} 
shows that both the QCD and electroweak two-loop corrections
give sizeable contributions of 10--15\% with respect to the one-loop result.

\begin{figure}[htb]
\centerline{
\psfig{figure=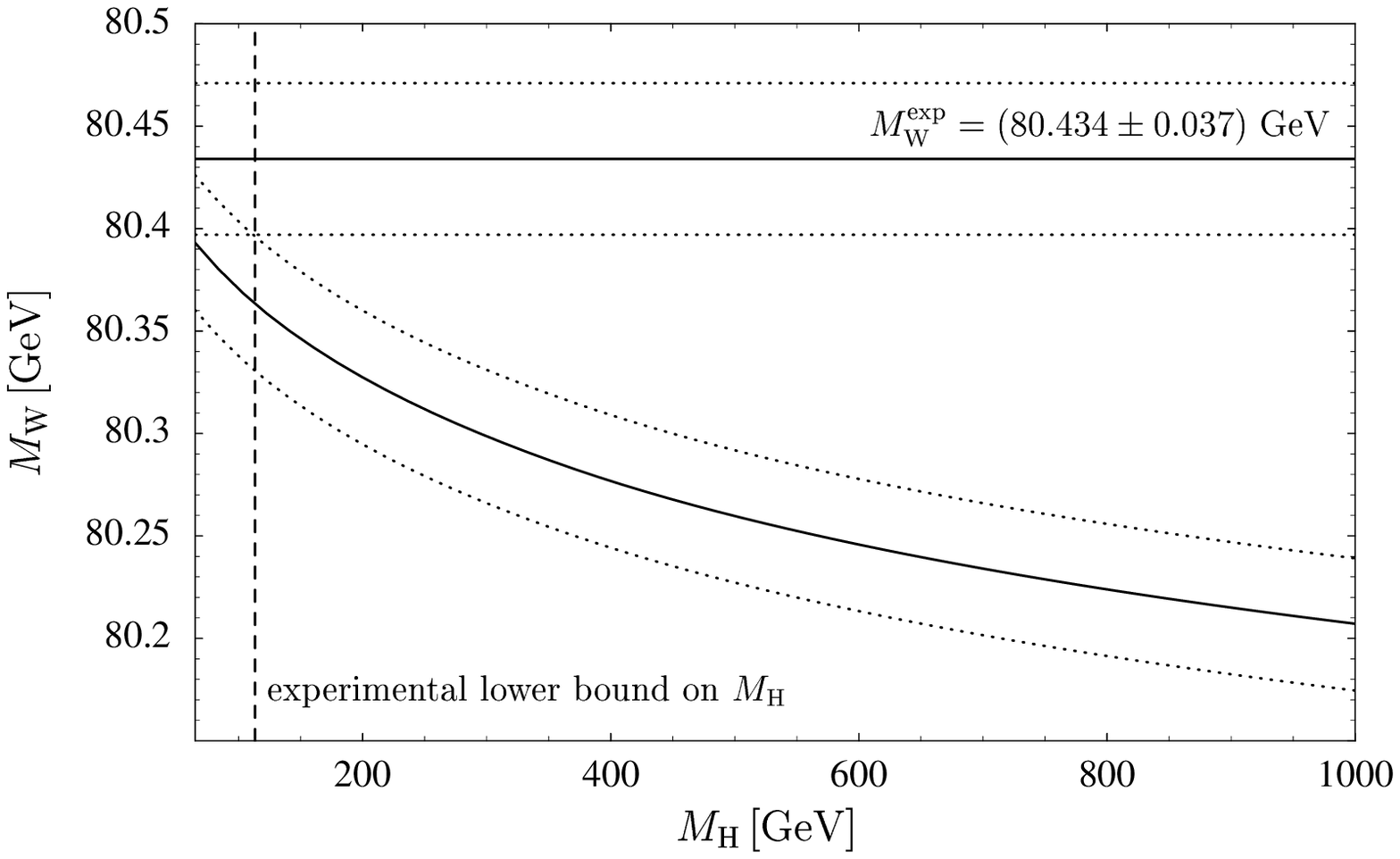,width=14cm}
}
\caption{{\small
The SM prediction for $\MW$ as a function of $\MH$ for 
$\mt = 174.3 \pm 5.1$~GeV is compared with the current experimental
value, $\MW^{\mathrm{exp}} = 80.434 \pm 0.037$~GeV~\cite{mori00}.
}}
\label{fig:MWpred}
\end{figure}

In Fig.~\ref{fig:MWpred} the prediction for $\MW$ derived from the result
\eqref{drtotal} and the relation \eqref{eq:delr} is compared with the
experimental value for $\MW$. Dotted lines indicate one standard deviation
bounds. The main uncertainties of the prediction originate from the
experimental errors of $\mt = (174.3 \pm 5.1)$ GeV \cite{C:pdg} and
$\Delta\alpha = 0.05954 \pm 0.00065$ \cite{eidjeg}. 
It is obvious that low
Higgs masses are favored; the new results on $\Delta r$ 
strengthen the tendency towards a lighter Higgs boson
(according to the following comparison).

These results can be compared with the results obtained by expansion of the
two-loop contributions up to next-to-leading order 
in $\mt$ \cite{ewmt2}. The predicted values
for $\MW$ for several values of $\MH$ are given in Tab.~\ref{tab:MWcomp}.
For the input parameters the values of \cite{ewmt2} have been
chosen, {\it i.e.}~$\mt = 175$~GeV, $\MZ = 91.1863$~GeV, $\De\al = 0.0594$,
$\alps(\MZ) = 0.118$.
\begin{table}[b]
\caption{Comparison between $\MW$--predictions from an NLO expansion in $\mt$
($\MW^{\stxt{expa}}$) and the full calculation ($\MW^{\stxt{full}}$). $\delta
\MW$ denotes the difference. }
\label{tab:MWcomp}
\setlength{\tabcolsep}{1.1pc}
\newlength{\digitwidth} \settowidth{\digitwidth}{\rm 0}
\catcode`?=\active \def?{\kern\digitwidth}
\begin{tabular*}{\columnwidth}{@{}r@{\extracolsep\fill}r@{\extracolsep\fill}
        r@{\extracolsep\fill}r@{\extracolsep\fill}r}
\hline
\rule{0mm}{1em}$\MH$ &
$\MW^{\stxt{expa}}$ &
$\MW^{\stxt{full}}$ &
$\delta \MW$ \\
\rule{0mm}{0mm} [GeV] &
[GeV] &
[GeV] &
[MeV] \\
\hline
$65$ & $80.4039$ & $80.3997$ & $4.2$  \\
$100$ & $80.3805$ & $80.3771$ & $3.4$ \\
$300$ & $80.3061$ & $80.3051$ & $1.0$ \\
$600$ & $80.2521$ & $80.2521$ & $0.0$ \\
$1000$& $80.2129$ & $80.2134$ & $-0.5$ \\
\hline
\end{tabular*}
\end{table}
Agreement is found between the results with maximal 
deviations of less than 5 MeV in $\MW$.
The deviations in the last column of Tab.~\ref{tab:MWcomp}
can of course not be attributed exclusively to differences in the 
two-loop top-quark and light-fermion
contributions, because the results also differ 
by a slightly different treatment of those higher-order terms that are
not yet under control, such as purely bosonic two-loop contributions,
and effects from scheme dependence.

Similar to \cite{DGPS}, a simple numerical
parametrization of our result for $\MW$ can be given by
the following expression:
\begin{equation}
\MW = \MW^0 - c_1 \, \mathrm{dH} - c_5 \, \mathrm{dH}^2 + c_6 \, \mathrm{dH}^4 
       - c_2 \, \mathrm{d}\al + c_3 \, \mathrm{dt} 
       - c_7 \, \mathrm{dH} \, \mathrm{dt} - c_4 \, \mathrm{d}\alps ,
\label{eq:simppar}
\end{equation}
where 
\[
\mathrm{dH} = \ln\left(\frac{\MH}{100 \gev}\right), \quad  
\mathrm{dt} = \left(\frac{\mt}{174.3 \gev}\right)^2 - 1, 
\]
\begin{equation}
\mathrm{d}\al = \frac{\De\al}{0.05924} - 1, \quad \quad
\mathrm{d}\alps = \frac{\alps(\MZ)}{0.119} - 1, 
\end{equation}
with the coefficients 
$\MW^0 = 80.3767$~GeV, $c_1 = 0.05613$, $c_2 = 1.081$, $c_3 = 0.5235$,
$c_4 = 0.0763$, $c_5 = 0.00936$, $c_6 = 0.000546$, $c_7 = 0.00573$.
and with $\MZ = 91.1785$ GeV.
The quality of the approximation~(\ref{eq:simppar}) to
our full result for $\MW$ is within $0.4$~MeV, 
allowing $\MH$ between 65 GeV and 1 TeV.


\section{Conclusion}

In this talk the realization of an exact two-loop calculation of fermionic
contributions in the full electroweak SM and its application to the
precise computation of $\Delta r$ has been reviewed. 
Numerical illustrations
were given for the results, which might serve as
ingredients for future SM fits.

\clearpage

\Acknowledgments
This work was supported by the Deutsche Forschungsgemeinschaft
(Forschergruppe ``Quantenfeldtheorie, Computeralgebra und
Monte Carlo Simulation'') and by the European Union
(HPRN-CT-2000-00149).

\end{document}